\documentstyle[12pt]{article}

\def \D {\hbox{d}}
\def \Log {\mathop{\rm Log}\nolimits}

\def \sinh{\mathop{\rm sinh}\nolimits}
\def \cosh{\mathop{\rm cosh}\nolimits}

\def\Metric{\sigma}
\def \bfE {{\bf E}}
\def \bfK {{\bf K}}
\def \bfR {{\bf R}}
\def \bfu {{\bf u}}

\begin{document}

\pagestyle{plain} 

\begin{center}
 {\bf NON-FUCHSIAN EXTENSION TO THE PAINLEV\'E TEST}
\end{center}

\vskip 0.5 truecm

{\bf Micheline Musette} and {\bf Robert Conte\dag}

\medskip

Dienst Theoretische Natuurkunde,
Vrije Universiteit Brussel,
\hfill \break \indent
B-1050 Brussel,
Belgique
\medskip

\dag
CEA, Service de physique de l'\'etat condens\'e,
Centre d'\'etudes de Saclay,
\hfill \break \indent
F-91190 Gif-sur-Yvette Cedex,
France

\vskip 0.5 truecm

\noindent {\it Short title}: Non-Fuchsian Painlev\'e test

\vglue 1.0truecm
\baselineskip=12truept

\noindent PACS: 0230, 0340 K

\vskip 0.8 truecm

\noindent {\it Keywords}
\par singularity analysis
\par Painlev\'e property

\vskip 0.8 truecm

{\it {\bf Abstract} --
We consider meromorphic particular solutions
of nonlinear ordinary differential equations
and perform a perturbation {\it \`a la} Poincar\'e
making their linearized equation non-Fuchsian at the movable pole
and Fuchsian at infinity.
When the nonlinear equation possesses movable logarithms,
they are detected sooner than with the perturbative (Fuchsian) Painlev\'e test.
}

\vfill
\noindent

\rightline{\noindent
           Phys.~Lett.~A, received 24 February 1995, revised 1 August 1995,
accepted 4 August 1995}
\rightline{\noindent \hfill Saclay SPEC 94/118}

\eject


\baselineskip=14truept 


\section{Introduction}
\indent

Nonlinear differential equations 
with fixed critical points
define a natural extension
\cite{PaiOeuvres} of linear equations.
Let us recall that a singular point is said
{\it critical} if several determinations of the solution are permuted
around it,
and {\it movable} (contrary {\it fixed}) if its location
depends on the initial conditions.
The {\it Painlev\'e property} (PP) of a differential equation,
is defined \cite{PaiOeuvres}
as the absence of movable critical points in the general solution
of the differential equation.

The {\it Painlev\'e test} \cite{Chamonix1994}
is the set of all methods
to build necessary conditions for a differential equation to possess the PP,
without guarantee on their sufficiency.
The most powerful of them is the ``$\alpha-$method'' of Painlev\'e
\cite{PaiBSMF},
but its difficulty is differential
since at each step one must integrate a differential equation.
All other methods
(Kowalevskaya \cite{Kowalevskaya1889,ARS1980}, 
 Bureau       \cite{Bureau1939},
 Conte, Fordy and Pickering \cite{FP1991,CFP1993})
have only an algebraic difficulty
and therefore are easy to automatize.

But all those algebraic methods share a restriction,
which this Letter proposes to remove,
thus adding to the Painlev\'e test a new algebraic method
which very often allows to conclude to a failure more rapidly than with the old
algebraic methods.

In section \ref{sectionPerturbative},
we recall the lemma of Painlev\'e which establishes the perturbative
framework,
then we define our extension to the Painlev\'e test.
Section \ref{sectionChazy} is a pedagogical example
quite useful to explain the method.
Two examples are then considered,
one from mathematics section \ref{eqsectionBureau},
the second one from physics section \ref{sectionBianchi}.
Both exhibit movable logarithms
and provide a shorter proof of failure of the test than before.

\section{The perturbative framework
\label{sectionPerturbative}}
\indent

All the methods of the Painlev\'e test \cite{Chamonix1994},
whether differential (Painlev\'e)
or algebraic (Kowalevskaya, Bureau, Conte {\it et al.}),
are based on the following lemma
of Painlev\'e \cite{PaiBSMF}.

{\it Lemma}.
Consider an algebraic differential equation of order $N$
\begin{equation}
\bfE(x,\bfu,\varepsilon)=0,
\end{equation}
(the boldface characters mean multicomponent)
depending analytically on a small complex parameter $\varepsilon$,
defined in the canonical form of Cauchy
\begin{equation}
{\D \bfu \over \D x}=\bfK[x,\bfu,\varepsilon],\
   x \in {\cal C},\ \bfu \in {\cal C}^N,\ \varepsilon \in {\cal C},
\label{eqLemma}
\end{equation}
with $\bfK$ assumed holomorphic at $\varepsilon=0$.
If the general solution of (\ref{eqLemma})
has no movable critical points
for every $\varepsilon$ except maybe for $\varepsilon=0$,
then
\begin{itemize}
\item{}
the point $\varepsilon=0$ is no exception,
i.e.~the general solution has also no movable critical points there,
\item{}
every coefficient of the Taylor series of $\bfu$
\begin{equation}
   \bfu=\sum_{n=0}^{+ \infty} \varepsilon^n \bfu^{(n)}
\label{eqPerturbu}
\end{equation}
has no movable critical points.
\end{itemize}

A detailed proof of this lemma has been established by
Bureau \cite{Bureau1939,BureauI},
in place of the too short proof by Painlev\'e,
who merely considered it as an immediate consequence of the classical
theorem of perturbations of Poincar\'e.

Let us introduce the notation
\begin{equation}
\bfE(x,\bfu,\varepsilon)
    \equiv\sum_{n=0}^{+ \infty} \varepsilon^n
 \bfE^{(n)}(x,\bfu^{(0)},\dots,\bfu^{(n)})=0
\end{equation}
in which the equation $\bfE^{(0)}(x,\bfu^{(0)}) = 0$ is nonlinear
and every equation
\hfill \par \noindent
$\bfE^{(n)}(x,\bfu^{(0)},\dots,\bfu^{(n)})=0,n\ge 1$
is linear inhomogeneous in $\bfu^{(n)}$. 

Consider an equation $\bfE(x,\bfu)=0$ independent of $\varepsilon$,
a case in which the lemma still applies (ref.~\cite{Goursat} vol.~III).
The equations
$\bfE(x,\bfu)=0$ and $\bfE^{(0)}(x,\bfu^{(0)}) = 0$
are then the same,
the equation $\bfE^{(1)}=0$ is the linearized of $\bfE$ at $\bfu^{(0)}$
and equations $\bfE^{(n)}=0, n \ge 2,$ only differ from $\bfE^{(1)}=0$
in the r.h.s.,
\begin{eqnarray}
n=0:\
\bfE^{(0)}
& \equiv &
\bfE(x,\bfu^{(0)}) = 0,
\label{eqNL0}
\\
n=1:\
\bfE^{(1)}
& \equiv &
\bfE'(x,\bfu^{(0)}) \bfu^{(1)} = 0,
\label{eqLin0}
\\
\forall n \ge 2:\
\bfE^{(n)}
& \equiv &
\bfE'(x,\bfu^{(0)}) \bfu^{(n)}
 + \bfR^{(n)}(x,\bfu^{(0)},\dots,\bfu^{(n-1)}) = 0.
\label{eqOrdern}
\end{eqnarray}

Suppose one knows a solution $\bfu^{(0)}$ which is
global and without movable critical points,
but which depends on an insufficient number $M<N$ of arbitrary parameters.
We require,
first,
that its perturbation (\ref{eqPerturbu})
in the neighborhood of $\varepsilon=0$
represents the general solution,
secondly,
that each $\bfu^{(n)}$, in particular $\bfu^{(1)}$ in this Letter,
be without movable critical points.
Painlev\'e showed (ref.~\cite{PaiBSMF} p.~209 note 1) that $M$ of the $N$
solutions of the linearized equation (\ref{eqLin0}) are already known,
these are the derivatives of $\bfu^{(0)}$ with respect to its $M$ parameters,
evidently without movable critical points.

Satisfying the first point is not so evident.
Indeed,
a peculiarity of nonlinear differential equations is that they sometimes
possess,
in addition to a general solution
and its particular solutions,
other solutions called singular solutions.
Those are impossible to obtain from the general solution
by giving special values
to the arbitrary integration constants.
For instance, the equation \cite{ChazyThese}
\begin{equation}
 u'''=2 u' u'' + 2 i u'' \sqrt{u'' - u'^2 - 1}
\end{equation}
has for general solution:
$u=e^{2 c_1 x + c_2} + (c_1^2 - 1) x /(4 c_1) + c_3$
and for singular solution:
$u=- \Log \cos (x + C_1) + C_2$,
showing the absence of correlation \cite{ChazyThese}
between the structure of singularities of the general solution
and that of the possible singular solutions.

Singular solutions are for sure excluded,
as requested by the definition of the PP
and the application of the above lemma,
if the equation $\bfE^{(1)}=0$ is exactly of order $N$.
Indeed, the singular solutions,
which generalize the notion of envelope,
satisfy a differential equation with an order smaller than $N$.

But the mean used by the available algebraic methods
to satisfy this requirement is to ask that the equation
(\ref{eqLin0}) possesses exactly $N$ Fuchs indices
[
for the basic vocabulary
(singular regular, singular irregular, Fuchsian, non-Fuchsian,
essential singularity, Fuchs indices)
and methods concerning linear differential equations
in the complex plane,
the reader can refer to the classical book of Ince \cite{Ince},
chapters XV to XVII
]
at the movable poles $x_0$ of $\bfu^{(0)}$ located at a finite distance,
which implicitly
discards possible non-Fuchsian solutions,
even if they have no movable critical points.

Let us briefly recall the differences between Fuchsian and non-Fuchsian
for our purpose which is the question of local singlevaluedness.

Near a regular singular point $x=0$,
there exist $N$ linearly independent solutions
\begin{equation}
x^{\lambda_i} \sum_{j=0}^{m_i} \varphi_{ij}(x) (\Log x)^{j},\
i=1,N
\label{eqFundamentalSetFuchs}
\end{equation}
in which the $\lambda_i$'s are complex numbers (the Fuchs indices),
$m_i$ positive integers (their multiplicity),
$\varphi_{ij}$ converging Laurent series of $x$ with finite principal parts.
The necessary and sufficient condition of local singlevaluedness of the
general solution of the linear equation is: $\lambda_i$ all integer,
no $\Log$ terms.

Near an irregular singular point $x=0$,
there exist $N$ linearly independent solutions
\begin{equation}
e^{Q_i(1/z_i)} x^{s_i} \sum_{j=0}^{m_i} \varphi_{ij}(z_i) (\Log x)^{j},\
z_i=x^{1/q_i},\
i=1,N
\label{eqFundamentalSetNonFuchs}
\end{equation}
in which
$q_i$ is a positive integer,
$Q_i$ a polynomial,
$s_i$ a complex number,
$\varphi_{ij}$ a {\it formal} Laurent series with finite principal part.
The question of local singlevaluedness of the general solution
cannot be settled so easily, because formal series are
generically divergent.

In this Letter, we go back to the lemma of Painlev\'e
and we implement two features.

The {\it first feature} is to also accept that $x_0$ be non-Fuchsian for
the linearized equation (\ref{eqLin0}):
we require the existence of a fundamental set of
solutions $\bfu^{(1)}$ of $\bfE^{(1)}=0$
which are locally single valued
near $\chi=x-x_0=0$,
and the same property for
a particular solution $\bfu^{(n)}$ of each $\bfE^{(n)}=0,n \ge 2$.

The {\it second feature} needs an additional assumption,
namely that the given solution $\bfu^{(0)}$ be known globally
(in closed form).
The singular points of equation (\ref{eqLin0}) can be classified into
three types:
\begin{enumerate}
\item
the fixed singularities of equation (\ref{eqNL0})
   located at a finite distance,

\item
the movable poles $x_0,x_1,\dots$ of $\bfu^{(0)}$
   located at a finite distance,

\item
the point at infinity, which is fixed.

\end{enumerate}

All three types can be studied because $\bfu^{(0)}$ is known globally.
The first type of singularities
must not be studied because the PP allows any behaviour around them.
Each point $x_0,x_1,\dots$ of the second type must be studied,
with the same requirements than in the first feature.
As to the third type, namely the point at infinity,
it should generally not be tested for it is fixed.
However, in case there are no critical singularities of the first type
(fixed critical singularities at a finite distance),
one can also require the existence of a fundamental set of solutions
locally single valued near $\infty$;
in particular, it may happen the favourable event that,
while the point $x_0$ is non-Fuchsian, the point $\infty$ is Fuchsian
and allows to conclude about {\it global} singlevaluedness.

\medskip

Our addition to the Painlev\'e test can be detailed as follows.

\begin{enumerate}
\item
Assume a given solution $\bfu^{(0)}$
which is global and has at least one
movable pole at a finite distance denoted $x_0$.

\item
Require the linearized equation at $\bfu^{(0)}$ to have exactly order $N$,
so as to discard singular solutions.

\item
Near every movable pole $x_0$ of $\bfu^{(0)}$ located at a finite distance,
build a fundamental set of solutions $\bfu^{(1)}$
and require it to be locally single valued.

\item
In case of at least one non-Fuchsian point among $x_0,x_1,\dots$ at the
preceding step,
and if there is no fixed critical singularity at a finite distance,
build a fundamental set of solutions $\bfu^{(1)}$ near $\chi=\infty$
and require it to be locally single valued.

\item
At each higher perturbation order $n \ge 2$,
build similarly particular solutions $\bfu^{(n)}$
and require the same properties.

\end{enumerate}

We will call a {\it family} {\it Fuchsian} or {\it non-Fuchsian}
according to the nature (Fuchsian or non-Fuchsian)
of the singular point $\chi=0$
of the linearized equation.

The most likely event to occur in our extension of the test,
leading to its failure,
is the detection of a movable logarithm
coming from two integer Fuchs indices at the point $\chi=\infty$,
of course under the condition of absence of any fixed critical singularity
at a finite distance.

The information (pass or fail) which we propose to extract is
\begin{enumerate}
\item
not accessible to the method of Kowalevskaya,
which accepts the Fuchsian families 
as well as the non-Fuchsian ones    
but does not consider the perturbation (\ref{eqPerturbu}),

\item
not accessible to the method of Bureau,
which rejects the non-Fuchsian families
(because the representation by divergent series forbids to conclude)
and does not consider the perturbation (\ref{eqPerturbu}),

\item
accessible to the method of Conte, Fordy and Pickering,
which rejects the non-Fuchsian families for the same reason than Bureau
and extracts the information from Fuchsian families
only at the expense of a {\it Fuchsian} perturbative computation
which may sometimes be long.

\end{enumerate}

\section{An explanatory example: Chazy's class III ($N=3,M=2$)
\label{sectionChazy}}
\indent

The simplified (i.e.~scaled) equation of class III of Chazy \cite{ChazyThese}
\begin{equation}
E(x,u) \equiv u''' - 2 u u'' + 3 u'^2 = 0.
\label{eqChazy}
\end{equation}
admits the global two-parameter solution
\begin{eqnarray}
u^{(0)}
& = &
c \chi^{-2} - 6 \chi^{-1},\ \chi=x-x_0,\ (x_0,c) \hbox{ arbitrary}.
\label{eqChazyOrder0}
\end{eqnarray}
For that equation, this solution arises from the {\it local} search
for all the families of movable singularities
$ u \sim u_0 \chi^p,\ E \sim E_0 \chi^q,\ \chi=x-x_0 \to 0$
represented by Laurent series with a finite principal part
\begin{eqnarray}
p=-1,\
q=-4,\
u^{(0)}
& = &
-6 \chi^{-1},
\label{eqChazyFamily1}
\\
p=-2,\
q=-6,\
u^{(0)}
& = &
c \chi^{-2} - 6 \chi^{-1},\ c \hbox{ arbitrary}.
\label{eqChazyFamily2}
\end{eqnarray}

The linearized equation at (\ref{eqChazyOrder0})
\begin{equation}
E'(x,u^{(0)}) u^{(1)} \equiv
 [              \partial_x^3
  - 2 u^{(0)}   \partial_x^2
  + 6 u^{(0)}_x \partial_x
  - 2 u^{(0)}_{xx}] u^{(1)} = 0,
\label{eqChazyLin}
\end{equation}
has effectively order $N$
(which means that solution (\ref{eqChazyOrder0}) is a particular one,
not a singular one),
it possesses the two single valued global solutions
$\partial_{x_0} u^{(0)}$ and $\partial_c u^{(0)}$,
i.e.~$u^{(1)}=\chi^{-3},\chi^{-2}$,
and it has only two singular points $\chi=0,\infty$.
The point $\chi=0$ is irregular singular of rank two
for the non-Fuchsian family (\ref{eqChazyFamily2})
(and regular singular for the Fuchsian family (\ref{eqChazyFamily1}),
 with Fuchs indices $-4,-3,-2$),
and it defines a third {\it formal} solution
admitting an essential singularity at $\chi=0$
(ref.~\cite{Ince} chap.~XVII)
\begin{equation}
\chi \to 0,\
c \not=0:\
u^{(1)}=e^{a / \chi} \chi^{s} w(\chi),
\end{equation}
in which
     $(a,s)$ denote constants and
     $w(\chi)$ a formal Taylor series generically asymptotic,
i.e.~with a null radius of convergence.
This is not the case here,
where computation yields $a=-2c,s=-2,w=1$,
thus defining the fundamental set of {\it global} solutions
\begin{eqnarray}
\forall \chi\
\forall c:\
u^{(1)} & = & \chi^{-2},\
              \chi^{-3},\
              (e^{-2 c / \chi} - 1 + 2 c \chi^{-1}) \chi^{-2} / (2 c^2)
\label{eqChazySGOrder1}
\end{eqnarray}
and solving the question for the perturbation order $n=1.$

The point $\chi=\infty$ is therefore, in this too simple an example,
useless to study.
This point is regular singular with Fuchs indices $2,3,4$,
and without any more computation one concludes to the {\it local}
singlevaluedness
since index $4$ cannot generate a logarithm.

{\it Remarks}.
\begin{itemize}
\item
Going on with the formalism of Painlev\'e's lemma at higher orders
constitutes the rigorous mathematical framework of the local representation
of the general solution obtained by Joshi and Kruskal \cite{JoshiKruskal1993}
\begin{equation}
u= - 6 \chi^{-1}
+ c \chi^{-2} (1 + z  - z^2 / 8  + z^3 / 144 - 7 z^4 / 13824
+ O(\varepsilon^5)),\
z=(\varepsilon / c) e^{- 2 c / \chi}.
\end{equation}
This representation reduces to the one given by Chazy (Taylor series in
$1/ \chi$)
if one starts from the Fuchsian family (\ref{eqChazyFamily1}).

\item
The lowering by $M=2$ units of the order of the linearized equation
is obtained by the change of function
\begin{equation}
u^{(1)}=\chi^{-3} v:\
(\partial_x + 3 \chi^{-1} - 2 c \chi^{-2}) v''=0,
\end{equation}
which yields the third global solution by three quadratures.

\end{itemize}

The simplified equation (\ref{eqChazy}),
which possesses the PP \cite{ChazyThese}
and therefore for which no $u^{(n)}$ is multivalued,
only shows the method.

We now illustrate on other examples the interest of non-Fuchsian families
to detect the presence of a movable critical singularity,
very often as soon as the first perturbation order.

\section{An equation of Bureau ($N=4,M=2$)
\label{eqsectionBureau}}
\indent

The equation (ref.~\cite{BureauMII} p.~79)
\begin{equation}
E(x,u) \equiv u'''' + 3 u u'' - 4 u'^2 = 0
\end{equation}
possesses the global two-parameter solution \cite{CFP1993}
\begin{equation}
u^{(0)}=c \chi^{-3} - 60 \chi^{-2},\ \chi=x-x_0,
\end{equation}
for which the linearized equation
\begin{equation}
E^{(1)} = E'(x,u^{(0)}) u^{(1)} \equiv
 [              \partial_x^4
  + 3 u^{(0)}   \partial_x^2
  - 8 u^{(0)}_x \partial_x
  + 3 u^{(0)}_{xx}] u^{(1)} = 0,
\label{eqBureauLin}
\end{equation}
whose only two singular points are $\chi=0$ and $\chi=\infty$,
admits the two global single valued solutions
$\partial_{x_0} u^{(0)}$ and $\partial_c u^{(0)}$,
i.e.~$u^{(1)}=\chi^{-4},\chi^{-3}$,
leaving only two other solutions to examine.

For $c=0$ it has four global single valued solutions
$u^{(1)}=\chi^{-5},\chi^{-4},\chi^{-3},\chi^{18}$,
and from now on we assume $c\not=0$.

The point $\chi=0$ is singular irregular with rank one,
and the two other solutions are non-Fuchsian and {\it formally} given as
\begin{equation}
\chi \to 0,\
c \not=0:\
u^{(1)}=e^{\pm \sqrt{-12c/ \chi}}\chi^{31/4} (1 + O(\sqrt{\chi})),
\end{equation}
detecting the presence in (\ref{eqBureauLin})
of an essential singularity at $\chi=0$,
but the generically null radius of convergence of the formal series forbids
to immediately conclude to the multivaluedness of $u^{(1)}$.

The point $\chi=\infty$ is singular regular, with Fuchs indices
$-18,3,4,5$.
The highest index, $5$, yields a converging series
$u^{(1)}=(1/\chi)^{5} w_{5}(1/\chi)$,
locally single valued.
As for the test for the existence of a series
$u^{(1)}=(1/\chi)^{-18} w_{-18}(1/\chi)$,
it detects a logarithm arising from the index $3$
when one tries to solve the recursion relation for the twenty-second
coefficient of the series $w_{-18}$.

One thus concludes here as soon as order one,
to be compared with
order seven necessary to ref.~\cite{CFP1993},
after a computation practically untractable without a computer.

{\it Remark}.
The lowering by $M=2$ units of the order of the linearized equation
(\ref{eqBureauLin}) is obtained with
\begin{equation}
u^{(1)}=\chi^{-4} v:\
(\partial_x^2 -16 \chi^{-1} \partial_x +3 c \chi^{-3} - 60 \chi^{-2}) v'' = 0,
\end{equation}
and it yields the two other solutions in global form
\begin{eqnarray}
c \not=0:\
v''_1
& = &
\chi^{-3} {}_{0} F_{1} (24;-3c/\chi)
=
\chi^{17/2} J_{23}(\sqrt{12 c/\chi}),
\\
v''_2
& = &
\chi^{17/2} N_{23}(\sqrt{12 c/\chi}),
\end{eqnarray}
where the hypergeometric fonction ${}_{0} F_{1} (24;-3c/\chi)$
is single valued and possesses an isolated essential singularity at $\chi=0$,
while the fonction $N_{23}$ of Neumann is multivalued because of a
$\Log \chi$ term.

{\it Remark}.
The value $n=7$ of the perturbation order in $\varepsilon$
needed by ref.~\cite{CFP1993} is the root of the linear
equation
$n (i_{\rm min}-p) + (i_{\rm max}-p)=-1$,
with $p=-2,i_{\rm min}=-5,i_{\rm max}=18$,
linking the pole order $p$ in the Fuchsian case $c=0$,
the smallest and the greatest Fuchs indices.
It expresses the condition for the first occurence of a power $\chi^{-1}$,
leading by integration to a logarithm,
in the r.h.s.~$R^{(n)}$ of the linear inhomogeneous equation
(\ref{eqOrdern}),
r.h.s.~created by the nonlinear terms $3 u u'' - 4 u'^2 $.

\section{An example in cosmology: Bianchi IX ($N=6,M=4$)
\label{sectionBianchi}}
\indent

The Bianchi IX cosmological model in vacuum
\cite{LandauLifshitzTheorieChamps}
is ruled by the sixth order system
\begin{equation}
\Metric^2 (\Log A)'' = A^2 - (B-C)^2
\hbox{ and cyclically},\
\Metric^2= \pm 1,
\label{eqBianchi1}
\end{equation}
and it does not possess the PP \cite{CGR1994,LMC1994}.
Let us prove it with our method.

In the neighborhood of the global solution
depending on the four arbitrary parameters $(k_1,k_2,\tau_1,\tau_2)$
\cite{Taub}
($x$ is here denoted $\tau$)
\begin{equation}
A^{(0)}= \Metric {k_1 \over \sinh k_1 (\tau-\tau_1)},\
B^{(0)}=C^{(0)}
       = \Metric
 {k_2^2 \sinh k_1 (\tau-\tau_1) \over k_1 \sinh^2 k_2 (\tau-\tau_2)},
\label{eqTaub}
\end{equation}
the perturbation
\begin{eqnarray}
& &
A= A^{(0)} (1 + \varepsilon A^{(1)} + O(\varepsilon^2))
\hbox{ and cyclically}
\end{eqnarray}
generates a linearized system whose order is indeed equal to $N=6$
(which proves that solution (\ref{eqTaub}) is a particular solution,
not a singular solution),
and the lowering by $M=4$ units of its order
is obtained by the change of function
dictated by the symmetry of the system:
$P^{(1)}=B^{(1)}+C^{(1)},M^{(1)}=B^{(1)}-C^{(1)}$ \cite{LMC1994}
\begin{eqnarray}
& &
\Metric^2 A^{(1)''} - 2 A^{(0)^2} A^{(1)} = 0,\
\label{eqTaub1A}
\\
& &
\Metric^2 P^{(1)''} - 2 A^{(0)} B^{(0)} P^{(1)}
 = 4 (A^{(0)} B^{(0)} - A^{(0)^2}) A^{(1)},\
\label{eqTaub1P}
\\
& &
\Metric^2 M^{(1)''} + 2 (A^{(0)} B^{(0)} - 2 B^{(0)^2}) M^{(1)} = 0.
\label{eqTaub1M}
\end{eqnarray}
Indeed, the four single valued global solutions
\begin{equation}
(A^{(1)},P^{(1)})=\partial_c (\Log A^{(0)},\Log (B^{(0)}+C^{(0)})),\
c=k_1,k_2,\tau_1,\tau_2,
\end{equation}
are those of the equations (\ref{eqTaub1A})--(\ref{eqTaub1P}),

\begin{equation}
(A^{(1)},P^{(1)} + 2 A^{(1)})=
\left\{
\begin{array}{l}
(
(\tau - \tau_1) \coth k_1 (\tau - \tau_1) - 1/k_1
,
0
),
\\
(
0
,
(\tau - \tau_2) \coth k_2 (\tau - \tau_2) - 1/k_2
),
\\
(
\coth k_1 (\tau - \tau_1)
,
0
),
\\
(
0
,
\coth k_2 (\tau - \tau_2)
),
\end{array}
\right.
\nonumber
\end{equation}
and there only remains to study the equation (\ref{eqTaub1M}),
the singular points of which (modulo the period of $\sinh$) are
$\tau-\tau_2=0$ and $\tau=\infty$.

At $\tau-\tau_2=0$,
the equation (\ref{eqTaub1M})
genericaly possesses irregular singular points of rank two
since the coefficient $B^{(0)^2}$ has there quadruple poles.
Its two non-Fuchsian solutions are {\it formally} (ref.~\cite{Ince} chap.XVII)
\begin{equation}
\tau - \tau_2 \to 0:\
M^{(1)}=
e^{\alpha / (\tau - \tau_2)}
\sum_{k=0}^{+ \infty} \lambda_k (\tau-\tau_2)^{k+s},\
\lambda_0 \not=0,
\end{equation}
with \cite{LMC1994}
\begin{equation}
\alpha=  \pm 2 k_1^{-1} \sinh k_1 (\tau_2 - \tau_1),\
s     =1 \mp 2          \cosh k_1 (\tau_2 - \tau_1).
\end{equation}
The two generically irrational values for the exponents $s$
allow to conclude
only if the Taylor series $\lambda_k (\tau-\tau_2)^{k}$ can be summed,
which is the case at least for $k_1=k_2=0$
where the two solutions are globally known \cite{LMC1994}:
\begin{eqnarray}
k_1=k_2=0:\
& &
{\D^2 M^{(1)} \over \D t^2}
 + \left({2 \over t^2} - {4 (t-1)^2 \over t^4} \right) M^{(1)}=0,\
t={\tau - \tau_2 \over \tau_1 - \tau_2},
\nonumber
\\
& &
M^{(1)}=
e^{-2/t} t^{-1},\
e^{-2/t} t^{-1} \int^{1/t} z^{-4} e^{4 z} \D z,
\end{eqnarray}
The second  solution implies the presence of a logarithmic branch point
at $t=0$, or at $t=\infty$ as well.
This singularity persists for $(k_1,k_2)\not=(0,0)$
and this proves the absence of the Painlev\'e property
for the Bianchi IX model in vacuum.


\section{Conclusion
\label{sectionConclusion}}
\indent

The main application of this extension to the Painlev\'e test
is the case of a non-Fuchsian family:
if the unperturbed solution is known in closed form
and if the linearized equation has no fixed singularity at a finite distance,
the study of the point at infinity
often allows to conclude to non-integrability
(in the sense of absence of the PP)
much more rapidly than the above mentioned algebraic methods.
The present method constitutes an algorithmic extension to the Painlev\'e test.

For the two examples given where the Painlev\'e test fails,
this failure was already known,
and we do not know yet if there exist equations where a failure would only be
detectable by the present extension and, of course, by the $\alpha-$method.

Nevertheless, the algorithmic cost of the present method is much smaller than
the cost of earlier methods,
and this is particularly sensitive on the example of Bureau.

Moreover, in our two examples, the perturbation order $n=1$ is sufficient
to dectect a failure,
which the perturbative Fuchsian test detects at the respective orders
$n=7$ (Bureau) and $n=2$ (Bianchi IX).
Despite the examination of several other equations without the PP
and admitting a non-Fuchsian family,
we could not find an example
requiring an order $n$ greater than one to exhibit a failure of the test.
Such a feature may be generic and certainly deserves future investigations.

{\it Acknowledgements}.
We thank D.~Bessis, A.~Latifi, A.~Mezincescu and A.~Pickering
for fruitful discussions.



\vfill \eject

\end{document}